# Mining The Successful Binary Combinations: Methodology and A Simple Case Study

**Yuval Cohen[1,2]**

**[1] Department of Industrial Engineering, The Open University of Israel,
Raanana, 43107, Israel**

**[2] Department of Industrial Engineering, Afeka Tel-Aviv Academic College of Engineering,
Tel-Aviv, 69107, Israel**

## Abstract

The importance of finding the characteristics leading to either a success or a failure is one of the driving forces of data mining. The various application areas of finding success/failure factors cover vast variety of areas such as credit risk evaluation and granting loans, micro array analysis, health factors and health risk factors, and parameter combinations leading to a product success. This paper presents a new approach for making inferences about dichotomous data. The objective is to determine rules that lead to a certain result. The method consists of four phases: in the first phase, the data is processed into a binary format of a truth table, in the second phase; rules are found by utilizing an algorithm that minimizes Boolean functions. In the third phase the rules are checked and filtered. In the fourth phase, simple rules that involve one to two features are revealed.

***Keywords:*** *Data Mining, Project Success, Rule Extraction, Knowledge Acquisition, Heuristics, Binary Data, Type dichotomy.*

## 1. Introduction

The Rules of success and failure as well as characterization of parameters that lead to desired results have immense importance in today's business world. For example, finding the underlying rules for credit-risk evaluation, insurance-application evaluation, and project evaluation are all very important management science problems [1, 2]. Though great advances had been achieved in this area, the search for techniques for finding such rules is as fervent as ever. The need to extract knowledge from data has spawned increasing efforts in trying to infer rules from databases [3]. For example, in the last two decades, the fields of neural networks and data mining have grown considerably.

This paper is taking a step forward in this direction. It offers a new approach in dealing with dichotomous data (fields with one of two values). Specifically, it characterizes combinations of features that lead to one of two results we call a success or a failure.

The paper presents a technique for rule extraction with three major constraints that differentiate it from the general data mining techniques:
1. All the data fields must be made dichotomous (0/1 values)
2. The population of records is classified into two groups: Success=1, and Failure=0.
3. The rules always associate records with the group defined as 1 (the Success group).

The presented approach finds the rules characterizing the desired combinations and expresses these rules in the most efficient way. In the business world, various phenomena could be classified as either success or failure. For example, a success could be when a customer in a supermarket purchases a bottle of wine, when an entrepreneur gets a loan, or when corporate sales grow over 50%. In other cases, phenomena could be classified into two values with no clear winner. For example, a population classified into patients under 21 years and patients over 21 years. The Success group in this latter case is chosen based on our interest in the group, or arbitrarily.

The question of interest is: "what are the rules that lead to a success or a failure?" Or in different words: what variables are associated with such success?

For example, if the purpose is to characterize customers that purchase wine, the question is what variables are associated with them. The variables could be the customer's gender, the customer's age, other products that were purchased, the time of day, the total bill amount in dollars, etc.

Notice however, that while gender or a product purchase is a 0/1 variable by nature, other variables are not. Thus, we have to segment the other variables into 0/1 groups by determining a threshold. For example, customers under 30 years of age and over 30 years, or bills under $ 100 and over $ 100.

Let us look at another case where a patient clinic has to allocate time slots for patient callers who schedule appointments with the doctors. The appointments could be either 15 minutes or 30 minutes appointments.

The clinic management is interested in categorizing the patients into two groups, asking the patients several yes/no questions should help determine whether to allocate 15 minutes or 30 minutes. Dichotomous variables could be whether the patient is smoking or not, whether body temperature is at least 2 degrees above normal, whether the patient is over 60 or not, etc.

Note that in such cases, the combination of values has a crucial importance. For example, the combination of a smoking person that is over 60 years old could trigger classification into a 30 minute appointment with the Dr., whereas a young smoking person and an old non-smoking person could be allocated 15 minutes appointments each.

Since we constrained ourselves in this paper to binary variables, the rules could be expressed as combinations (or strings) of ones and zeros of the corresponding variables. For example, if we are trying to characterize beer consumers, using their smoking habits, age, and gender, it is reasonable to define the rule:

X= smoking/non (smoking=1, non=0).

Y =gender (male=1, female=0),

Z=age over or under 30 (under=1, over=0).

So if we found that all the smoking males under 30 in the sample are beer consumers, it could be expressed as (X=1, Y=1, Z=1) or 111 or simply XYZ. When a rule contains a zero we use the NOT=' sign (e.g., X NOT=X'). For example, if we try to characterize healthy people we may find that all the non-smokers under 30 are healthy (X=0 and Z=1) expressed as X'Z. So X'Z is the rule that has been found.

When there are several rules of "success" the rules are expressed using the "+" sign serving as Boolean "or". For example, if it is found that all organic food buyers are either women or non-smokers, using the classification above, the organic food buyers are characterized as either X=0, or Y=0, or simply X' + Y'. This expression is the rule we were looking for.

However, the way to the desired rule, is the really important part. This part is described through a four-phase mechanism that identifies rules from binary data. The rest of the paper is arranged as follows: Section 2 discusses some related literature; Section 3 describes the phases in subsections 3.1 to 3.4. In section 4 we provide a case study, and use it in sub sections 4.1 to 4.4 to illustrate the four phases respectively.

## 2. Related Literature

One of the learning techniques that generates a set of rules for integer and binary data (but still needs extensive set of training examples) is the technique of forming a decision tree, [4, 5, 6]. The ID3 and C4.5 algorithms by Quinlan [7, 8] serve as good examples of learning algorithms that are suitable for building rules.

However, Quinlan's algorithm needs large amounts of data for the learning process and cannot cope with bad or missing data. A very critical view of the above methods appears in [9].

Some implementations of classifying and characterizing desired combinations of attributes is shown in the literature. Credit-risk evaluation for granting loans based on the client characteristics is dealt in [1, 2, 10]. Viaene et. al. [11] are dealing with classifying customers for insurance fraud detection.

The Logit model [12,13,14,15,16] is the most common and well known regression based approach for discrete and binary data. While it cannot deal with the effect of combinations, it does find the effect of each single independent variable (the main effects). Logit differs from regular regression by handling data in which the dependent variable is binary or even discrete ordinal and the independent variables can be either continuous or categorical. The main idea is to find a relationship between predetermined values of independent variables and the probability that the dependent variable is a success (or that it is a failure). Logit model utilizes a regression procedure and maximum likelihood principle, to estimate the main effect of each independent variable. However, the Logit model can not deal with the effect of combinations of variables and so, is not suitable for the case studied in this paper.

In this paper and in some of the previous methods forming Winning rules is based on the "general principle of inductive learning often called Ockham's razor: The most likely hypothesis is the simplest that is consistent with all observations." [5] p. 534. An Ockham algorithm is "an algorithm that is capable of finding consistent hypothsis that achieves a significant compression of the data it represents" [5] p. 560. Ockham algorithms are further discussed in [17]. This paper utilizes a unified algorithm of Quine [18, 19] and McCluskey [20, 21] that is an Ockham algorithm. The algorithm is discussed and explained thoroughly in chapter 4 of Kohavi [22]. This algorithm is a generalization of Karnaugh maps devised by Karnaugh [23] for small problems. It is not surprising that the map method of Karnaugh is also an Ockham algorithm.

## 3. Proposed Methodology

The methodology of the proposed technique is divided into four phases:

1. Get the data and construct a Truth Table
2. Form Winning Rules - Using algorithms by Quine [18, 19] and McCluskey [20, 21]
3. Filter outliers and non-efficient elements
4. Check for main effects and effects of all pairs of combinations.

The first phase involves getting the data inverting the non-binary data into binary data constructing the Truth Table. Getting the data may involve sampling. The Truth Table tells us what combinations have proven successful, and what combinations have been failures.

The second phase forms rules by describing the Truth table as compactly as possible. This is done by adjusting an algorithm by Quine [18, 19] and McCluskey [20, 21] that minimizes logical functions in general.

The third phase, is implementing a filter based on the Pareto principle to eliminate inefficient rules from phase 2.

The fourth phase, checks for rules that cover many instances, but may have been rejected in step 2 due to exceptions or lack of data.

Each of the phases would be discussed in more details below.

### 3.1 First phase: Truth Table Construction

The purpose of this step is to classify the observed data records into one of two groups (success or failure). If the data is coming from a mechanical or digital system where the same inputs always result in the same output the first phase is fairly simple:

Phase I for a fully determined system
    Stage 1: Get the data
    Stage 2: Process the data into dichotomous values (using a threshold when necessary).
    Stage 3: Construct a truth table

Missing combinations in the data are spots of uncertainty and in typical conservative treatment should be considered as losing combinations (just to be on the safe side). Table 1 is an illustration of a very simple truth table.

The described process in Table 1 is deterministic, but in many cases the process or the system may be stochastic. For example consider purchase of a product, human reaction, and most business phenomena – for each process/phenomenon - repeating the same input may result in different output reaction.

Table 1. Truth table for the rule:

(if X=0 and Y=0 Then Z=1) or in short: X'Y'

| Input | | Output |
|:---:|:---:|:---:|
| **X** | **Y** | **Z** |
| 0 | 0 | 1 |
| 0 | 1 | 0 |
| 1 | 0 | 0 |
| 1 | 1 | 0 |

When such a non-deterministic system is involved, some preparatory stage must precede the construction of truth table, and the procedure is altered as follows:

Phase I for a system with uncertain response
    Step 1: Get the data
    Step 2: Process the data into dichotomous values (using a threshold
               when necessary )
    Step 3: Construct a frequency table with the following fields: Binary combination, # successes, # failures, and
    % successes: # successes / (# successes + # failures)
    Step 4: Classify Winning binary combinations of data using threshold values of:
    "% successes" and "# successes".
    Step 5: Construct truth table

### 3.2 Second Phase: Minimizing The Boolean Function

The truth table holds every combination that was classified as a success. For $k$ dichotomous attributes, the truth table has $2k$ entries. This is a very long and inefficient way of describing the rules of success or failure. Moreover, it grows exponentially. We need better rules that will capture the common features of large groups of combinations. Like other methods, the proposed method is based on Ockham's razor philosophy: The most likely hypothesis is the simplest that is consistent with all observations." (See section 2: Related Literature.) Thus, an Ockham type algorithm by Quine [18, 19] and McCluskey [20, 21] is adopted by the authors. This algorithm comes from the realm of digital design and when applied to the truth table it generates minimal Boolean function of that table. Minimal Boolean function of a truth table is a function that describes the table using minimal number of terms. Minimizing Boolean functions is an Ockham algorithm since minimal Boolean function is by definition the simplest function. Minimizing Boolean functions include the pursuit of powerful rules with fewest variables as possible. This could be explained in a more intuitive way as follows:

Suppose for example, that we seek combinations of characteristics that are associated with lung patients that developed cancer. Furthermore, suppose that all these combinations contain cigarette smokers and there are only very few smokers without cancer. Classifying smoking as the rule for cancer is the most efficient rule in this case. Smoking is one variable. Rules based on a single variable are the most efficient way to describe large groups. In general rules with pair of variables are less efficient then rules with one variable since they cover fewer success combinations. Rules with three variables cover less combinations then rules with two variables, and so on. In general, the rule is stronger and more efficient when it contains fewer variables.

In the early days of digital design, logical variables were constructed using physical gates and vacuum bulbs. Digital designers were struggling to minimize the cost of representing a truth table. The only way to do this was to minimize logical functions. While small functions with up to 5 variables were minimized using the prevalent Karnaugh maps [23], the algorithm for the general case is much less known and combines two algorithms: the first by Quine [18, 19] and the second by McCluskey [20, 21] into one framework. The framework and the details of the algorithm along with examples and discussion are presented in Kohavi [22].

### 3.3  Third Phase: Filtering rules

The result of phase two is a set of rules that exactly matches the original truth table. While this is much more compact then a truth table, it still may have quite a few specific rules for isolated cases. Like most real world systems, the Pareto rule holds in this case. Namely, 20% of the rules describe 80% of the success cases. Moreover, the desirability of keeping a rule that describe less then 1% of the cases is usually low enough to ignore this rule altogether. This brings us to the third phase where non desirable rules are filtered out.

All we have to do for the third phase is to set a level of rule acceptance. Say we chose the rule acceptance level to be 5%, than any rule that does not describe at least 5% of the cases is filtered out.

While the third phase filters out inefficient rules, it may happen that a very efficient rule with only one or two variables has not been revealed so far. For this to happen it is enough that one combination (or more) belonging to the rule is missing, or was not recorded by a mistake. These cases are treated in the fourth phase.

### 3.4  Fourth Phase: Checking for Main and Secondary Effects

Main effects are rules of a single variable (that is, a single attribute). Effects of pair of attributes are secondary effects.

Many efficient rules could be missing due to missing data and exceptions. The fourth phase is checking for such cases and amends them. This phase is relatively simple. Checking for main effects requires only that we collect for each variable: (1) the % of successes when the variable is one, and (2) the % when the variable is zero. Exceeding a predetermined percentage (e.g., 95% successes when the variable is 1) would inaugurate a new rule (with the predetermined of at least 95% accuracy). We also check for every pair of variables and their success/failure percentage. Note that the complexity of checking combinations grows immensely and that the power of the resultant rules drops significantly with the growing number of attributes. It is therefore that we do not recommend going beyond pairs for this brute force enumeration.

## 4.  Case Study

The purpose of the case study is to illustrate the stages of the proposed methodology. Due to the obvious space constraints, a small case study is chosen. Thus, a major advantage of the third phase (dealing with many attributes) is bypassed. However, the case study illustrates all the other important points.

In this case study we are trying to characterize the type of project managers that lead their project to a success.

### 4.1  Phase I

However, Between the indicators that characterize human personalities it is very convenient for our case study to adopt the model by Gustav Jung [24] based on four dichotomies. These four dichotomies are answers to the four questions: (1) Where do you focus your attention? (2) In what way do you take in information? (3) In what way do you make decisions? (4) Hoe do you deal with the outer world? [25]

The answers to these questions are described on dichotomous scale:

1. **Extraversion** vs. **Introversion**
2. **Sensing** vs. **Intuition**
3. **Thinking** vs. **Feeling**
4. **Judging** vs. **Perceiving**

We shall use the following definitions:
E=1 if a candidate is Extravert, and zero otherwise.
S=1 if a candidate is Sensing, and zero otherwise.
T=1 if a candidate is Thinking, and zero otherwise
J=1 if a candidate Judging, and zero otherwise.

Phase I aims at constructing the truth table. Since we have $k$=4 dichotomous attributes we have 24=16 possible combinations. The case study database consisted of records of successfully completed projects of a large consulting (undisclosed) company and their managers. Successful project is defined as one that was completed on time, on budget and within specifications. The managers were asked to take a short Myers Briggs test to reveal their four dichotomies. For each combination of traits, the number of successful projects is counted and percentage is calculated in Table 2. Next step is to extract rules (classify the "successful" binary combinations) using threshold values of "% successes" and "# successes".

In case of total randomness, each of the 16 combinations would average 1/16 of the observations or close to 6.25% of the cases. Therefore, the threshold should be set higher than that. In the case study we set the threshold at 7% (or 70 observations). Setting the threshold is somewhat arbitrary decision based on the analyst discretion. However, in many cases the Pareto rule may work, where we set the threshold to differentiate between the higher 20% (or so) and the rest of the pack.

The following truth table (Table 3) is calculated by simply converting all values in the right side of Table 2 that are under the threshold to zero and those above the threshold to one (remember that the threshold is 7%).

## 4.2 Phase II: Extracting Rules

The aim of this stage is to form rules based on the truth table. We applied the method by Quine [18, 19] and McCluskey [20, 21] to get the following rules: (The method is explained thoroughly in Kohavi [22].) Project managers of successful consulting projects are characterized by the following combinations:

1) Extravert, Thinking
2) Extravert, Sensing, Perceiving

The mathematical notation for the rules is based on initial letters of the dichotomous fields: For example, E=1 means extraversion, and E=0 means introversion.

This may be written for short as E=Extraversion, and for the other fields: S=Sensing, T=Thinking, J=Judging. Also, we use ∪ for Boolean "or" and ∩ for Boolean "and".

Table 2: Success frequency table for all combinations

| Binary combination | | | | # success Number of Success-ful Projects | % |
|---|---|---|---|---|---|
| Extravert Introvert 1=E 0=I | Sensing iNtuition 1=S 0=N | Thinking Feeling 1=T 0=F | Judging Perceive 1=J 0=P | | |
| 0 | 0 | 0 | 0 | 2 | 0.2% |
| 0 | 0 | 0 | 1 | 1 | 0.1% |
| 0 | 0 | 1 | 0 | 6 | 0.6% |
| 0 | 0 | 1 | 1 | 16 | 1.6% |
| 0 | 1 | 0 | 0 | 5 | 0.5% |
| 0 | 1 | 0 | 1 | 27 | 2.7% |
| 0 | 1 | 1 | 0 | 32 | 3.2% |
| 0 | 1 | 1 | 1 | 31 | 3.1% |
| 1 | 0 | 0 | 0 | 10 | 1.0% |
| 1 | 0 | 0 | 1 | 36 | 3.6% |
| 1 | 0 | 1 | 0 | 102 | 10% |
| 1 | 0 | 1 | 1 | 180 | 18% |
| 1 | 1 | 0 | 0 | 40 | 4% |
| 1 | 1 | 0 | 1 | 140 | 14% |
| 1 | 1 | 1 | 0 | 202 | 20% |
| 1 | 1 | 1 | 1 | 170 | 17% |
| Total 8 | 8 | 8 | 8 | 1000 | 100% |

Writing the letter means that its value is one, adding the apostrophe (') to the letter means that its value is zero. For example, "E" to represents E=1, adding 'the apostrophe (') as in "E' " represents E=0. Thus, the rule is:

Y = (E∩T) ∪ (E∩S∩J).

It is customary to replace ∪ by the "+" sign and treat ∩ as a multiplication resulting in:

$$Y = (E \cap T) \cup (E \cap S \cap J) = ET + ESJ \qquad (1)$$

Table 3: Truth Table for Table 2 (with Threshold of 7%)

| Binary combination | | | | Result |
|---|---|---|---|---|
| Extravert Introvert | Sensing iNtuition | Thinking Feeling | Judging Perceive | 1=success 0=Failure |
| 1=E 0=I | 1=S 0=N | 1=T 0=F | 1=J 0=P | |
| 0 | 0 | 0 | 0 | 0 |
| 0 | 0 | 0 | 1 | 0 |
| 0 | 0 | 1 | 0 | 0 |
| 0 | 0 | 1 | 1 | 0 |
| 0 | 1 | 0 | 0 | 0 |
| 0 | 1 | 0 | 1 | 0 |
| 0 | 1 | 1 | 0 | 0 |
| 0 | 1 | 1 | 1 | 0 |
| 1 | 0 | 0 | 0 | 0 |
| 1 | 0 | 0 | 1 | 0 |
| 1 | 0 | 1 | 0 | 1 |
| 1 | 0 | 1 | 1 | 1 |
| 1 | 1 | 0 | 0 | 0 |
| 1 | 1 | 0 | 1 | 1 |
| 1 | 1 | 1 | 0 | 1 |
| 1 | 1 | 1 | 1 | 1 |

Since this is a relatively small problem (with k<6) the same results of the Quine & McCluskey algorithm could be achieved and verified graphically using a Karnaugh map (see figure 1). Karnaugh map is a graphical tool for minimizing Boolean functions. Karnaugh map is composed of a matrix, in which each entry corresponds to a single combination. The value of each entry is binary (1 or 0, for success or failure).

Forming the matrix starts with splitting the binary attributes of the problem into two separate groups: a variable group for the columns and the rest of the attributes for the rows. For example, in figure 1, four binary attributes: E, S, T, J are divided into E, S for the columns, and T, J for the rows. Each column corresponds to a single combination of values of E and S (one of: 00,01,10,11).

Each row in figure 1 corresponds to a single combination of values of T and J (one of: 00,01,10,11). Y=ET+ESJ =
(1) Extravert, Thinking Or
(2) Extravert, Sensing, Judging.

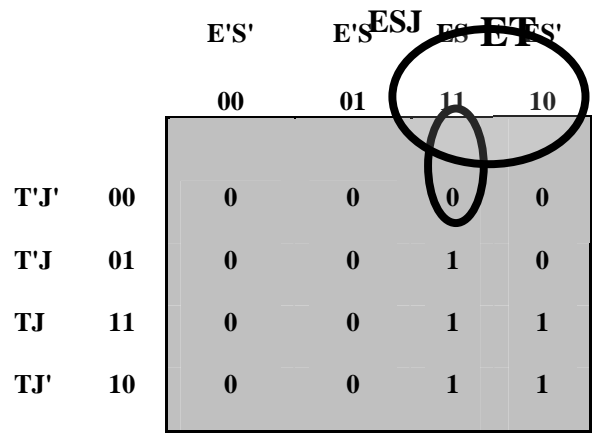

Fig. 1 Karnaugh map for the case study. The circles correspond to the rules.

The value of each entry in the Karnaugh map (in figure 1) is the result (success/failure) of the value combination formed by the column and the row. For example, the upper left corner has the first column corresponding to E'S'=00, and first row corresponding to T'J'=00. So the upper left corner entry corresponds to E'S'T'J'=0000. Its value is zero corresponding to a failure. As another example, the lower right corner corresponds to ES'TJ'=1010 and its value "1" represents a success. The column and row state combinations are ordered in the map so that between any pair of neighboring columns (or rows) there would be only one change of one field (bit).

For example, between (0,0) and (0,1) there is one bit change. An example for an illegal adjacency is (0,1) and (1,0) since there are two bit changes between the two.

Circling the largest groups of "1" that cover all the "1" and nothing but the "1" give the desired rules. For example, circling the four "1" on the bottom- right corner, corresponds to the common values of all entries in this group: in this case E=1, T=1. (E∩T). The circled groups (rules) can only have 2 or 4 or 8 entries and fully contained circles are dominated by the bigger group which is also the more general rule.

It is therefore, that Karnaugh maps could handle only up to 5 of attributes (for more on Karnaugh maps see Kohavi (1978)). Note that the method of Quine and McCluskey minimizes the binary function in the general case for any number of attributes.

## 4.3 Phase III: Filtering Out Rules

The filtering procedure is described using the following procedure:

I. Make a list of the rules, compute their corresponding percentage of the original success cases:

A. original successes:

$$P(ET) = 10\% + 18\% + 20\% + 17\% = 65\%, \qquad (2)$$

$$P(ESJ) = 14\% + 17\% = 31\% \qquad (3)$$

$$Total\_Successes = P(ET) + P(ESJ) - P(ESTJ) = $$
$$AC + ABC'D = 65\% + 31\% - 17\% = 79\% \qquad (4)$$

B. Percent of the original successes:

$$ET: \ 65/79 = 82\% \qquad (5)$$

$$ESJ: \ 31/79 = 39\% \qquad (6)$$

II. Sort the list by the percentage (The case study is too small, so it is sorted already).

III. Set the coverage level - the percentage of successes that you like to describe with rules. For illustration, we shall describe two cases: (1) coverage of 80% and (2) 90%.

IV. Follow the steps of the loop below:
   A) From the remaining rules, choose the rule with greatest percentage.
   B) Increase the "Success Coverage" by the chosen rule percentage.
   C) Reduce the percentage of the remaining rules by their overlap with the chosen group of rules.
   D) If the "Success Coverage" exceeds the threshold from step III - Stop;
   Otherwise delete the chosen rule from the list of remaining rules and go to A.

For example:
• The ET rule satisfies the threshold of 85% (it covers 82%>80% of the successes)
• The ET rule is not enough for the 90% threshold (82%<90%). We choose the next rule ESJ and we reach a coverage of 100% of the successes (which must be satisfactory).

## 4.4 Phase IV: Main and Secondary Effects

To In this phase we revisit sequentially the percentage of cases for each attribute or pair of attributes and decide which of them may have been ignored due to minor inconsistencies that can be tolerated. The case study is used for illustrating the fourth phase.

A. Revisiting the main effects

Since there are four attributes in the case study there are eight rules of single attribute to consider: E, E, S, S', T, T', J and J' (explicitly: E=1, E=0, S=1, S=0, T=1, T=0, J=1, J=0 ).

For example, consider the rule E meaning E=1 (Extravert) in our case study. From Table 2 this rule covers: 878 cases (87.8% of the population). However, this rule is inconsistent with the following entries in Tables 3: ES'TJ', EST'J', ES'T'J. So using the rule E=1 (Extravert) we have the following probability of error (using Table 1, and "#" to replace the word "number"):

P(Added Error) =
       = (Deviation from expected #)/(expected #)

Deviation = (Expected #)-(ES'TJ', EST'J', ES'T'J cases)
Expected # = (3 cases)(total/2k)) = 3*(1000/16)=187 cases.
ES'TJ', EST'J', ES'T'J cases = 86 cases
So,
P(Added Error) = (187-86)/187 = 54%. (7)

The overall error is the added error multiplied by its weight: (3/8)*(0.54)+(5/8)*(0)= 20%

In general the probability of error has to be weighed against the simplicity it brings (i.e. the number of rules it saves). In this case it saves one rule by replacing the two rules: ET, ESJ. So we have to weigh adding 20% error against saving one rule. These computations and decision repeat for all 8 potential rules.

B. Revisiting effects of all attribute pairs

Since there are four binary attributes (E,S,T,J) in the case study we have to consider the following 24 pair combinations: ES, E'S, ES', E'S', ET, E'T ET', E'T', EJ, E'J, EJ', E'J', ST, S'T, ST', S'T', SJ, S'J, SJ', S'J', TJ, T'J, TJ', T'J'.

ET is already part of the rules. For each pair, the computations are analogous to the computations of the single attribute.

For example, consider the rule EJ for the case study. Like all rules of attribute pairs it has four combinations:
1. ESTJ, meaning E=1,S=1,T=1,J=1 -included in Table 3
2. ES'TJ, meaning E=1,S=0,T=1,J=1 -included in Table 3
3. EST'J, meaning E=1,S=1,T=0,J=1 -included in Table 3
4. ES'T'J, meaning E=1,S=0,T=0,J=1 -Not included

For the added ES'T'J, the probability rule of added error is (using "#" to replace the word "number"):

P(Added Error) =
          =(Deviation from expected #)/(expected #)
Expected # = (total/2k)) = (1000/16) = 62 cases.
Deviation = (Expected #)-(ES'T'J cases)=
= 62-36 = 26 cases
P(Added Error) = 26/62 = 42%.                    (8)

The overall error is the added error multiplied by its weight: $(1/4)*(0.42)+(3/4)*(0) = 10.5\%$

If rule EJ replaces ESJ the only gain from the replacement is the ability to ignore attribute S (Sensing). The decision maker have to decide whether to ignore Sensing and have 10.5% error probability, or to eliminate this error probability by considering the Sensing.

## 5. Complexity as a Function of the Number of Attributes

Let us define $k$ as the number of attributes in the problem. As the problem becomes bigger, $k$ grows and the consequences are as follows:

- In phase 1: The number of combinations (rows in the truth table) is $2k$. This is an exponential growth.
- In phase 2: If $k>5$ Karnaugh map can no longer describe it. Instead, the Quine and McCluskey algorithm must be applied. However, the complexity of Quine and McCluskey algorithm grows exponentially with k.
- In phase 3: As $k$ grows the number of rules grows considerably, and filtering out rules becomes more of an issue. When it comes to human decision, too many rules complicate things, and we may be willing to trade the exactness of describing successes for simplicity.
- In phase 4: The number of single attribute computations is $2*k$ (each of the $k$ attributes can be either 1 or 0). The number of attribute pair computations is the multiplication of all the value combinations of the pair (22) by the number of pairs:
  $(k-1)+(k-2)+\dots1 = ((k-1)*k)/2$.
  For example, in 4.4.2 above, the number of pairs is:
  $(22)*((k-1)*k)/2 = 4*((4-1)*4)/2 = 4*12/2 = 24$ (9)

Overall, the number of computations is proportional to $k^2$ ($O(k^2)$). However, each single computation directly depends on the number of combinations and thus, grows exponentially with $k$.

## 6. Conclusions

This paper presents an approach for finding the binary combinations leading to a specified result. The approach is based on four phases and utilizes the fact that some data could easily be transformed into binary data as done in the first phase. In the second phase, we minimize the Boolean function. The third phase filters out superfluous outliers of the second phase, and the fourth phase appends missing combinations missing from the second phase. While the example in this paper is small, the method is very efficient with much larger systems.

## References

[1] B. Baesens, R. Setino, C. Mues, J. Vanthienen, "Using Neural Network Rule Extraction and decision tables for credit-risk evaluation", Management Science, Vol. 49, No. 3, 2003, pp. 312-329.

[2] A. Steenackers, M. J. Goovaerts, "A credit scoring model for personal loan, Insurance", Math Economics, Vol. 8, 1989, pp. 31-34.

[3] R. Andrews, J. Diederich, A. B. Tickle, "A survey and critique of techniques for extracting rules from trained neural networks", Knowledge Based Systems, Vol. 8, No. 6, 1995, pp. 373-389.

[4] T. Mitchell, "Decision tree learning", in T. Mitchell, Machine Learning, McGraw-Hill, 1997, pp. 52-78.

[5] S. Russell, P. Norvig, J. F. Canny, J. Malik, and D. D. Edwards, Artificial intelligence a modern approach, (Prentice Hall Series in Artificial Intelligence). Englewood Cliffs, NJ: Prentice Hall, 1995.

[6] P. Winston, "Learning by building identification trees", in P. Winston, Artificial Intelligence, Addison-Wesley Publishing Company, 1992, pp. 423-442.

[7] J. R. Quinlan, "Induction of decision trees", Machine Learning, Vol. 1, 1986, pp. 81-106.

[8] J. R., Quinlan, C4.5 programs for machine learning, Chambery, France: Morgan Kaufman, 1993.

[9] K. Chelst, "Can't See the Forest Because of the Decision Trees: A Critique of Decision Analysis in Survey Texts", Interfaces, Vol. 28, No. 2, 1998, pp. 80-98.

[10] N. Capon, Credit scoring systems, a critical analysis, Journal of Marketing, Vol. 46, 1982, pp. 82-91.

[11] S. Viaene, D. A. Derrig, B. Baesens, G. Dedene, A comparison of state of the art classification techniques for the auto-mobile insurance claim fraud detection, The Journal of Risk and Insurance, Vol. 69, No. 3, 2002, pp. 373-421.

[12] P. D. Allison, Logistic Regression Using the SAS System: Theory and Application, Cary, NC: SAS Institute, 1991.

[13] P. D. Allison, Survival Analysis Using the SAS System: A Practical Guide. 1995, Cary, NC: SAS Institute

[14] W. H. Greene, Econometric Analysis (Fourth edition). Prentice Hall,2000.

[15] S. J. Long, and J. Freese, Regression Models for Categorical Dependent Variables Using STATA, College Station, TX: STATA Press, 2001.

[16] S. J. Long, Regression Models for Categorical and Limited Dependent Variables. Advanced Quantitative Techniques in the Social Sciences. Sage Publications, 1997.

[17] H. J. William, and O. J. Berger, Ockham's razor and Bayesian analysis, American Scientist. Vol. 80, No. 1, 1992, pp. 64-72.

[18] W. V. Quine, The problem of simplifying truth functions, American Mathematics Monthly, Vol. 59 No.8, 1952, pp. 521-531.

[19] W. V. Quine, A way to simplify truth functions. American Mathematics Monthly, Vol. 62 No.9, 1955, pp. 627-631.

[20] E. J McCluskey, Minimization of Boolean functions, Bell System Technical Journal, Vol. 35 No. 6, 1956, pp.1417-1444.

[21] E. J. McCluskey, and H. Schorr, "Essential multiple-output prime implicants in mathematical theory of automata", Proceedings of the Polytechnique Institute Brooklyn Symposium, Vol. 12, 1962, pp. 437-457.

[22] Z. Kohavi, Switching and Finite Automata Theory (2nd ed.), McGraw Hill, 1978.

[23] M. Karnaugh, "The map method for synthesis of combinatorial logic circuits", Transactions of AIEE, Vol. 72 No. 9, 1953, pp. 593-599.

[24] C. G. Jung, Psychological types (Collected works of C. G. Jung, volume 6). (3rd ed.). Princeton, NJ: Princeton University Press. (Translated from German), 1971.

[25] A. Furnham, "The big five versus the big four: the relationship between the Myers-Briggs Type Indicator (MBTI) and NEO-PI five factor model of personality", Personality and Individual Differences, Vol. 21, No. 2, 1996, pp. 303-307.

**Yuval Cohen** received: PhD in Industrial Engineering (IE) from the University of Pittsburgh in 1998, M.Sc. in IE 1992 from the Technion (IIT), and a BSc. In IE in 1988 from Ben-Gurion University.; He worked as an Industrial Engineer at Tefen-USA during 1988 while finishing his PhD. He was a senior operations analyst at FedEx Ground 1988-2002, A senior lecturer at the open University of Israel 2002-Current; And A senior lecturer at Afeka Tel-Aviv college of engineering since 2009. He wrote many papers on various IE subjects. He is interested in operations research and data mining and is a member of INFORMS and IIE.

.